\documentclass[12pt]{iopart}
\usepackage{iopams}  
\usepackage{setstack}

\begin{document}
\title{On a method for constructing the Lax pairs for integrable models via quadratic ansatz}

\author{I T Habibullin$^{1,2}$, A R Khakimova$^{2}$}

\address{$^1$Institute of Mathematics, Ufa Scientific Center, Russian Academy of Sciences,
112, Chernyshevsky Street, Ufa 450008, Russian Federation}
\address{$^2$Bashkir State University, 32 Validy Street, Ufa 450076 , Russian Federation} 
\eads{\mailto{habibullinismagil@gmail.com}, \mailto{aigulya.khakimova@mail.ru}}

\begin{abstract}
A method for constructing the Lax pairs for nonlinear integrable models is suggested. First we look for a nonlinear invariant manifold to the linearization of the given equation. Examples show that such invariant manifold does exist and can effectively be found. Actually it is defined by a quadratic form. As a result we get a nonlinear Lax pair consisting of the linearized equation and the invariant manifold. Our second step consists of finding an appropriate change of the variables to linearize the found nonlinear Lax pair. The desired change of the variables is again defined by a quadratic form. The method is illustrated by the well-known KdV equation and the modified Volterra chain. New Lax pairs are found. The formal asymptotic expansions for their eigenfunctions are constructed around the singular values of the spectral parameter. By applying the method of the formal diagonalization to these Lax pairs the infinite series of the local conservation laws are obtained for the corresponding nonlinear models.

\end{abstract}


\maketitle

\eqnobysec

\section{Introduction}

Linearization of nonlinear equations at the vicinity of their arbitrary solutions is often used in the integrability theory. For instance, any evolutionary type symmetry
\begin{equation} \label{evol_sym}
	u_{\tau}=g(x,t,u,u_1,\dots,u_l), \quad u_j=D_x^ju
\end{equation}
of the partial differential equation
\begin{equation} \label{diff_eq}
	u_t=f(u,u_1,\dots,u_k)
\end{equation}
is effectively found as a solution $U=g(x,t,u,u_1,\dots,u_l)$ of the linearized equation
\begin{equation} \label{lin_eq}
	U_t=\left(\frac{\partial f}{\partial u}+\frac{\partial f}{\partial u_1}D_x+\frac{\partial f}{\partial u_2}D_x^2+\dots\frac{\partial f}{\partial u_k}D_x^k\right)U.
\end{equation}
Here $D_x$ is the operator of the total differentiation with respect to $x$.
The present article is devoted to further development of the approach to the problem of constructing the Lax pairs for integrable equations discussed in our recent article \cite{HKP} with M.N.Poptsova. In \cite{HKP} we suggested a method for constructing the recursion operator via linear invariant manifolds to their linearized equations. It provides a direct tool for finding the recursion operators and might be useful, for instance, in the case when the Lax pair to the equation is not known.

Recall the necessary definitions. Consider a surface in the space of the dynamical variables labeled by uppercase letters $\left\{U_j\right\}_0^\infty$ defined by an equation of the form 
\begin{equation} \label{surf_eq}
	H(U,U_1,\dots, U_m;u,u_1,\dots,u_{m_1})=0.
\end{equation}
Here the dynamical variables labeled by lowercase letters $u,u_1,u_2,\dots$ are considered as parameters. The surface (\ref{surf_eq}) is referred to as an invariant manifold to the linearized equation (\ref{lin_eq}) if the following condition 
\begin{equation} \label{cond}
	\left.D_tH\right|_{(1.2)-(1.4)}=0
\end{equation}
holds everywhere on the surface (\ref{surf_eq}) identically for all values of the variables $\left\{u_j\right\}$.
In a similar way the invariant manifolds are defined for the hyperbolic type PDE and the discrete models.
In our opinion the following two classes of the invariant manifolds worth attention:
\begin{itemize}
\item $H$ is a linear form $H=\sum^{m}_{j=1}\alpha_j(u,u_1,\dots)U_j$;
\item $H$ is a quadratic form $H=\sum^{m}_{i,j=1}\alpha_{ij}(u,u_1,\dots)U_iU_j$.
\end{itemize}
The former kind invariant surfaces are connected with the recursion operators while the latter can be used for finding the Lax pairs. Let us illustrate these two observations by the well known KdV equation
\begin{equation} \label{kdv_eq}
	u_t=u_3+uu_1.
\end{equation}
The linearized equation to (\ref{kdv_eq}) 
\begin{equation} \label{lin_kdv}
	U_t=U_3+uU_1+u_1U
\end{equation}
admits a linear invariant manifold of order $m=3$, i.e. corresponding to the linear form $H$
\begin{equation} \label{lin_invmn_kdv}
	H:=U_3-\frac{u_2}{u_1}U_2+\left(\frac{2u}{3}+\lambda\right)U_1-\left(\frac{2uu_2}{3u_1}-u_1+\frac{u_2}{u_1}\lambda\right)U=0.
\end{equation}
Where $\lambda\in C$ is an arbitrary parameter. Equation (\ref{lin_invmn_kdv}) can be rewritten as follows 
\begin{equation} \label{RU_lambdaU}
RU=\lambda U,
\end{equation}
where $R$ coincides with the well known recursion operator for the KdV equation: $R=D_x^2+\frac{2u}{3}+\frac{u_x}{3}D_x^{-1}$. Actually a couple of the equations (\ref{lin_kdv}), (\ref{RU_lambdaU}) can be regarded as a Lax pair for the equation (\ref{kdv_eq}). However this Lax pair differs from the classical one where instead of the third order equation we have the second order one. Actually system of equations (\ref{lin_kdv}), (\ref{RU_lambdaU}) is closely connected with the well-known system of equations the ``squares" of the eigenfunctions of the classical Lax pairs satisfy to.  Such kind Lax pairs have earlier been considered in \cite{IbragimovShabat}-\cite{Sul}. Note that equation (\ref{lin_kdv}) does not admit any linear invariant manifold of the order $m=2$. In the present article we observe that it is possible to assign to (\ref{kdv_eq}) the second order invariant manifold defined by a quadratic form. Let us first slightly simplify equation (\ref{lin_kdv}) by the potentiation $U=W_x$ which leads to 
\begin{equation} \label{lin_W_kdv}
W_t=W_3+uW_1.
\end{equation}
In \S2 we show that equation (\ref{lin_W_kdv}) admits the second order quadratic invariant manifold of the form 
\begin{equation} \label{invmn_W_kdv}
W_2W-\frac{1}{2}W_1^2+\left(\frac{u}{3}-\lambda\right)W^2=0.
\end{equation}
Note that formula (\ref{invmn_W_kdv}) appears in the frame of the finite-gap integration (see, for instance, formula (16) in \cite{Adler}). It can be verified that equations (\ref{lin_W_kdv}), (\ref{invmn_W_kdv}) constitute a nonlinear Lax pair for the KdV equation (\ref{kdv_eq}). The point transformation $W=\varphi^2$ reduces (\ref{lin_W_kdv}), (\ref{invmn_W_kdv}) to the usual Lax pair.

The article is organized as follows. In \S 2 we explain the method of constructing the Lax pairs with examples of the KdV equation and the Volterra lattice. Then in \S 3 we apply the method to two equations of the KdV type found in \cite{SvS} the Lax pair for which were not known before. In the last section \S 4 we demonstrate that the found systems are true Lax pairs since they admit formal eigenfunctions which allow one to produce the infinite series   of the conservation laws. 

\section{Illustrative examples of the application of the algorithm for searching the Lax pair.}

In this section we explain with some examples the algorithm of constructing the Lax pairs via quadratic forms.

\subsection{ Korteweg-de Vries equation}
As the first example we consider the KdV equation (\ref{kdv_eq}). For the technical reasons we have to simplify the linearized equation (\ref{lin_kdv}) by the following differential substitution $U=W_x$ by keeping the variables $u, u_x, u_{xx},...$ unchanged. Then obviously (\ref{lin_kdv}) is transformed into 
\begin{equation} \label{lin_Wt_kdv}
W_t=W_{xxx}+uW_x.
\end{equation}
Observe that the coefficients of (\ref{lin_Wt_kdv}) depend on $u$ only, while the coefficients of (\ref{lin_kdv}) depend on $u,u_x$. Therefore we can look for the invariant manifold to (\ref{lin_Wt_kdv}) depending on the parameter $u$ only: 
\begin{equation} \label{quadr_linmn_W2_kdv}
W_{xx}=F(W,W_x,u).
\end{equation}
Here we do not impose any additional assumption on the equation (\ref{quadr_linmn_W2_kdv}). The unknown F is found from the following equation derived as the consistency condition of the equations (\ref{lin_Wt_kdv}) and (\ref{quadr_linmn_W2_kdv})
\begin{equation} \label{eq_found_F_kdv}
\left.\frac{d}{dt}F(W,W_x,u)-\frac{d^2}{dx^2}(W_{xxx}+uW_x)\right|_{(1.6),(2.1),(2.2)}=0.
\end{equation}
After evaluating the derivatives and simplifying due to the equations (\ref{kdv_eq}), (\ref{lin_Wt_kdv}) and (\ref{quadr_linmn_W2_kdv}) we get:
\begin{eqnarray} \label{eq_found_F_compl_kdv}
\fl 3F_{uu}u_xu_{xx}+(3W_xF_{Wu}+W_x+3F_{W_xu}F)u_{xx}+(3F_{Wuu}W_x+3F_{W_xuu}F+3F_{W_xu}F_u)u_x^2 \nonumber\\
\fl +F_{uuu}u_x^3+(3F_{W_x}F_{W_xu}F+3F_{WWu}W_x^2-F_{W_x}W_x+3W_xF_{W_xu}F_W+6W_xF_{W_xWu}F\nonumber\\
\fl +3W_xF_{W_xW}F_u+3F_{Wu}F+2F+3FF_{W_xW_x}F_u+3F_{W_xW_xu}F^2)u_x+F_{WWW}W_x^3\\
\fl +F_{W_xW_xW_x}F^3+3F_{W_xW}F^2+3W_xF_{W_x}F_{W_xW}F+3W_xFF_{W_xW_x}F_W+3F_{W_x}F^2F_{W_xW_x}\nonumber\\
\fl +3W_x^2F_{W_xWW}F+3W_x^2F_{W_xW}F_W+3W_xF_{WW}F+3W_xF_{WxWxW}F^2=0.\nonumber
\end{eqnarray}
By collecting the coefficients before the product $u_xu_{xx}$ we find the following equation
\begin{equation*} \label{found_F_eq1_kdv}
\frac{\partial^2}{\partial u^2}F(W,W_x,u)=0
\end{equation*}
which implies
\begin{equation} \label{F_via_F1_F2_kdv}
F(W,W_x,u)=F_1(W,W_x)u+F_2(W,W_x).
\end{equation}
By means of (\ref{F_via_F1_F2_kdv}) the coefficient before $u_{xx}u$ in (\ref{eq_found_F_compl_kdv}) gives
\begin{equation*} \label{eq_found_F1_1_kdv}
F_1(W,W_x)\frac{\partial}{\partial W_x}F_1(W,W_x)=0.
\end{equation*}
Since the relation $F_1(W,W_x)=0$ leads to a contradiction we put $F_1(W,W_x)=F_1(W)$. Now we continue the investigation of the equation (\ref{eq_found_F_compl_kdv}) by using the specifications of $F$ obtained above. By collecting the coefficients before $u_{xx}$ we find 
\begin{equation*} \label{eq_found_F1_2_kdv}
3\frac{\partial}{\partial W}F_1(W)+1=0
\end{equation*}
and therefore 
\begin{equation} \label{eq_F1_kdv}
F_1(W)=-\frac{1}{3}W+C_1.
\end{equation}
Let us compare the coefficients of the term $uu_x$:
\begin{equation*} \label{eq_found_F2_1_kdv}
\frac{\partial^2}{\partial W_x^2}F_2(W_x,W)=\frac{1}{W-3C_1}.
\end{equation*}
The latter gives rise to
\begin{equation} \label{eq_F2_kdv}
F_2(W_x,W)=\frac{W_x^2}{2(W-3C_1)}+F_3(W)W_x+F_4(W).
\end{equation}
Let us substitute the specifications (\ref{eq_F1_kdv}), (\ref{eq_F2_kdv}) again into (\ref{eq_found_F_compl_kdv}) and collect the coefficients before $u_x$:
\begin{equation*} \label{eq_found_F3_kdv}
\frac{\partial}{\partial W}F_3(W)+\frac{1}{W-3C_1}F_3(W)=0.
\end{equation*}
Solution to this equation is 
\begin{equation} \label{eq_F3_kdv}
F_3(W)=\frac{C_2}{W-3C_1}.
\end{equation}
The coefficient before $uW_x$ gives $C_2=0$. The coefficient before $u$ generates the second order ODE for the last unknown function $F_4(W)$ which is nothing else but the Euler equation:
\begin{equation*} \label{eq_found_F4_kdv}
\frac{d^2}{dW^2}F_4(W)+\frac{1}{W-3C_1}\frac{d}{dW}F_4(W)-\frac{1}{(W-3C_1)^2}F_4(W)=0.
\end{equation*}
The equation is immediately solved
\begin{equation} \label{eq_F4_kdv}
F_4(W)=\frac{C_3}{W-3C_1}+C_4(W-3C_1).
\end{equation}
Summarizing the reasonings above we find
\begin{equation} \label{eq_F_kdv}
F(W,W_x,u)=(C_1-\frac{1}{3}W)u+\frac{W_x^2}{2W-6C_1}+\frac{C_3}{W-3C_1}+C_4(W-3C_1).
\end{equation}
The constant parameter $C_1$ is easily removed by the transformation $W\rightarrow \tilde{W}=W-3C_1,$ hence we can take $C_1=0$. Then by putting $C_3=0$ and $\lambda :=C_4$ we get $F=-\frac{1}{3}uW+\lambda W+\frac{W_x^2}{2W}$. Then equations (\ref{lin_Wt_kdv}), (\ref{quadr_linmn_W2_kdv}) can be written as follows:
\begin{equation} 
\label{Lax_W_kdv}
\left \{
\begin{array}{l}
W_{xx}=\left(\lambda-\frac{1}{3}u\right)W+\frac{W_x^2}{2W},\\
W_t=\left(2\lambda+\frac{1}{3}u\right)W_x-\frac{1}{3}u_xW.
\end{array} 
\right.
\end{equation}
Now it is evident that invariant surface (\ref{quadr_linmn_W2_kdv}) is defined by the quadratic form $H=W_{xx}W-\left(\lambda-\frac{u}{3}\right)W^2-\frac{1}{2}W_x^2$.
Note that the point transformation $W=\varphi^2$ reduces (\ref{Lax_W_kdv}) to the usual Lax pair for the KdV equation
\begin{equation} 
\label{Lax_kdv}
\left \{
\begin{array}{l}
\varphi_{xx}=\left(\frac{1}{2}\lambda-\frac{1}{6}u\right)\varphi,\\
\varphi_t=\left(2\lambda+\frac{1}{3}u\right)\varphi_x-\frac{1}{6}u_x\varphi.
\end{array} 
\right.
\end{equation}

\subsection{Modified Volterra chain}
As the second illustrative example we take the well known modified Volterra chain 
\begin{equation} \label{modVch}
p_t=-p^2(p_1-p_{-1}).
\end{equation}
Here the sought function $p$ depends on the discrete $n$ and continuous $t$: $p=p_n(t)$. For the simplicity we omit $n$ and use an abbreviated notation by writing $p_n=p$, $p_{n+1}=p_1$, $p_{n-1}=p_{-1}$ and so on.

Linearization of the equation (\ref{modVch})
\begin{equation} \label{lin_modVch}
P_t=-p^2(P_1-P_{-1})-2p(p_1-p_{-1})P
\end{equation}
contains an explicit dependance on the dynamical variables $p_1,p,p_{-1}$. In order to simplify the further necessary computations it is reasonable to find more convenient form of the linearized equation (\ref{lin_modVch}). By applying the substitution $P=-p^2(U_1-U_{-1})$ we reduce it to the following form 
\begin{equation} \label{lin_U_modVch}
U_t=-p^2(U_1-U_{-1}),
\end{equation} 
where the r.h.s. depends only on $p,U_1,U_{-1}$. Initiated by this circumstance we look for the invariant manifold to (\ref{lin_U_modVch}) in the form 
\begin{equation} \label{invmn_U1_F_modVch}
U_1=F(U,U_{-1},p).
\end{equation}
The unknown $F$ is found from the defining equation 
\begin{equation} \label{eq_found_FG_modVch}
\left.\frac{d}{dt}U_1-D_nU_t\right|_{(2.13),(2.15),(2.16)}=0.
\end{equation}
To exclude $U_{\pm2}$ we use the relations
\begin{equation} \label{Up2_Um2_modVch}
U_2=F(F(U,U_{-1},p),U,p_1)\quad \mbox{and} \quad U_{-2}=G(U,U_{-1},p_{-1})
\end{equation}
where G is found as the ``inverse" to $F$:
\begin{equation} \label{Um1_G_modVch}
U_{-1}=G(U_1,U,p).
\end{equation}
By evaluating the derivatives, shifting the arguments and substituting the   expressions for all of the involved variables in terms of the dynamical ones in (\ref{eq_found_FG_modVch}) we obtain a comparatively simple equation for the unknowns $F$ and $G$\begin{eqnarray} \label{eq_found_FG_compl_modVch}
\fl F_{U_{-1}}(U,U_{-1},p)\Bigl(G(U,U_{-1},p_{-1})-U\Bigr)p_{-1}^2+F_p(U,U_{-1},p)p^2p_{-1}-F_p(U,U_{-1},p)p^2p_1\nonumber\\
\fl +\Bigl(F(F(U,U_{-1},p),U,p_1)-U\Bigr)p_1^2-F_U(U,U_{-1},p)p^2\Bigl(F(U,U_{-1},p)-U_{-1}\Bigr)=0.
\end{eqnarray}
By applying the operator $\frac{\partial^3}{\partial p_{-1}^3}$ to (\ref{eq_found_FG_compl_modVch}) we find 
\begin{eqnarray} \label{eq1_found_FG_modVch}
\fl F_{U_{-1}}(U,U_{-1},p)\Bigl(p_{-1}^2\frac{\partial^3}{\partial p_{-1}^3}G(U,U_{-1},p_{-1})\nonumber\\+6p_{-1}\frac{\partial^2}{\partial p_{-1}^2}G(U,U_{-1},p_{-1})+6\frac{\partial}{\partial p_{-1}}G(U,U_{-1},p_{-1})\Bigr)=0.
\end{eqnarray}
Since the first factor doesn't vanish then (\ref{eq1_found_FG_modVch}) impies the Euler equation for $G$ which is easily solved:
\begin{equation*} \label{G_via_G1G2G3_modVch}
G(U,U_{-1},p_{-1})=G_1(U,U_{-1})+\frac{G_2(U,U_{-1})}{p_{-1}^2}+\frac{G_3(U,U_{-1})}{p_{-1}}.
\end{equation*}
Substitute the latter into (\ref{eq_found_FG_compl_modVch}) and afterward differentiate it twice with respect to $p_{-1}$ and find 
\begin{equation*} \label{G1_modVch}
G_1(U,U_{-1})=U.
\end{equation*}
Now we differentiate (\ref{eq_found_FG_compl_modVch}) three times with respect to $p_1$ and obtain again the Euler equation solution to which is of the form 
\begin{equation*} \label{F_via_F1F2F3_modVch}
F(U,U_{-1},p)=F_1(U,U_{-1})+\frac{F_2(U,U_{-1})}{p}+\frac{F_3(U,U_{-1})}{p^2}.
\end{equation*}
By applying $\frac{\partial^2}{\partial p_1^2}$ to (\ref{eq_found_FG_compl_modVch}) we find 
\begin{equation*} \label{F1_modVch}
F_1(U,U_{-1})=U_{-1}.
\end{equation*}
Then we substitute the simplification into (\ref{eq_found_FG_compl_modVch}) and differentiate the equation obtained with respect to $p_{-1}$. As a result we get the following three relations
\begin{eqnarray*} \label{eqs_found_F2F3G3_modVch}
\fl 1) \frac{\partial F_3(U,U_{-1})}{\partial U_{-1}}G_3(U,U_{-1})=0,\\
\fl 2) G_3(U,U_{-1})=F_2(U,U_{-1}),\\
\fl 3) F_3(U,U_{-1})=\frac{1}{2}\frac{\partial F_2(U,U_{-1})}{\partial U_{-1}}G_3(U,U_{-1}).
\end{eqnarray*}
Let us concentrate on the first one. It yields either $\frac{\partial F_3(U,U_{-1})}{\partial U_{-1}}=0$ or $G_3(U,U_{-1})=0$. The latter case implies due to 2) and 3) $F_2(U,U_{-1})=F_3(U,U_{-1})=0$ and leads to a negative result. The former case gives:
\begin{eqnarray*} \label{F2F3G3_modVch}
\fl 1) F_3(U,U_{-1})=F_3(U),\\
\fl 2) G_3(U,U_{-1})=F_2(U,U_{-1}),\\
\fl 3) F_2(U,U_{-1})=\sqrt{4F_3(U)U_{-1}+F_4(U)}.
\end{eqnarray*}
Due to these relations we can specify $F$ and $G$:
\begin{eqnarray} \label{Up1_modVch}
U_1=F(U,U_{-1},p)=U_{-1}+\frac{F_3(U)}{p^2}+\frac{\sqrt{4F_3(U)U_{-1}+F_4(U)}}{p},
\end{eqnarray}
\begin{eqnarray} \label{Um2_modVch}
U_{-2}=G(U,U_{-1},p_{-1})=U+\frac{G_2(U,U_{-1})}{p_{-1}^2}+\frac{\sqrt{4F_3(U)U_{-1}+F_4(U)}}{p_{-1}}.
\end{eqnarray}
Express $U_{-1}$ from the equation (\ref{Up1_modVch}) and then find $U_{-2}$ by applying the operator $D_n^{-1}$
\begin{eqnarray} \label{Dnm1Up1_modVch}
U_{-2}=U+\frac{F_3(U_{-1})}{p_{-1}^2}+\frac{\sqrt{4F_3(U_{-1})U+F_4(U_{-1})}}{p_{-1}}.
\end{eqnarray}
Comparison of (\ref{Um2_modVch}) with (\ref{Dnm1Up1_modVch}) leads to the equations:
\begin{eqnarray} \label{F3F4G2_modVch}
\fl 1) G_2(U,U_{-1})=F_3(U_{-1}),\nonumber \\
\fl 2) 4F_3(U)U_{-1}+F_4(U)=4F_3(U_{-1})U+F_4(U_{-1}).
\end{eqnarray}
Equation (\ref{F3F4G2_modVch}) convinces us that functions $F_3, F_4$ are linear:
\begin{equation*} \label{F3F4_modVch}
F_3(U)=c_1U+c_2, \quad F_4(U)=4c_2U+c_4.
\end{equation*}
Let us summarize the reasonings above and obtain the manifold desired 
\begin{eqnarray*} \label{Up1_F_modVch}
U_1=U_{-1}+\frac{c_1U+c_2}{p^2}+\frac{\sqrt{4(c_1U+c_2)U_{-1}+4c_2U+c_4}}{p}.
\end{eqnarray*}
By setting $c_2=c_4=0, c_1=\lambda$ we get
\begin{eqnarray} \label{invmn_final_modVch}
U_1=U_{-1}+\frac{\lambda}{p^2}U+\frac{2}{p}\sqrt{\lambda UU_{-1}}.
\end{eqnarray}
Since we have a square root then we change the variables in such a way $U=\varphi^2$ in order to get rid the root:
\begin{eqnarray} \label{U_varphi_modVch}
\varphi_1^2=\varphi_{-1}^2+\frac{\lambda}{p^2}\varphi^2+\frac{2\sqrt{\lambda}}{p}\varepsilon \varphi\varphi_{-1}, \quad \varepsilon^2=1.
\end{eqnarray}
By taking the square root from both sides in (\ref{U_varphi_modVch}) we find
\begin{eqnarray} \label{Lax_pair_varphi1_modVch}
\varphi_1=\frac{\sqrt{\lambda}}{p}\varphi +\varepsilon\varphi_{-1}.
\end{eqnarray}
By applying the same change of the variables $U=\varphi^2$ to the equation (\ref{lin_U_modVch}) we bring it to the form
\begin{eqnarray} \label{Lax_pair_varphit_modVch}
\varphi_t=-\frac{1}{2}\lambda\varphi +\varepsilon\sqrt{\lambda}p\varphi_{-1}.
\end{eqnarray}

The equations system (\ref{Lax_pair_varphi1_modVch}), (\ref{Lax_pair_varphit_modVch}) coincides with the well-known Lax pair for the modified Volterra lattice.

\section{New Lax pairs.}
In this section we search Lax pairs for the following two KdV type integrable models
\begin{eqnarray} \label{eq1_SvSok}
u_t=u_{xxx}+\frac{1}{2}u_x^3-\frac{3}{2}u_x\sin^2u,
\end{eqnarray}
\begin{eqnarray} \label{eq2_SvSok}
u_t=u_{xxx}-\frac{3u_xu_{xx}^2}{2(1+u_x^2)}+\frac{1}{2}u_x^3
\end{eqnarray}
found in \cite{SvS} as a result of the symmetry classification (see also \cite{SokolovMeshkov1}). A connection of these equations with the integrable hyperbolic type equations is discussed in \cite{SokolovMeshkov}. The recursion operators to (\ref{eq1_SvSok}), (\ref{eq2_SvSok}) have been found in \cite{{HKP}}. In the present article the Lax pairs to them are found via quadratic invariant manifolds to their linearizations.

\subsection{Equation (\ref{eq1_SvSok})}
Let us begin with the equation (\ref{eq1_SvSok}). First we evaluate the linearized equation
\begin{eqnarray} \label{eq1_lin_SvSok}
U_t=U_{xxx}+\frac{3}{2}(u^2_x-\sin^2 u) U_x-3u_x\sin u\cos uU.
\end{eqnarray}
Coefficients of the equation (\ref{eq1_lin_SvSok}) explicitly depend on $u$ and $u_x$. By direct computations we proved that there is no any differential substitution converting equation (\ref{eq1_lin_SvSok}) to a more simple linear equation with the coefficients depending only on $u$ or only on $u_x$. Let us look for a quadratic form $H=\Sigma^{2}_{i,j=0}\alpha_{ij}U_iU_j$ such that the surface 
\begin{eqnarray} \label{Heqs0_SvSok}
H=0
\end{eqnarray}
defines an invariant manifold for the equation (\ref{Heqs0_SvSok}). Since the coefficients in (\ref{eq1_lin_SvSok}) depend on $u,u_x$ we assume that the coefficients $\alpha_{ij}$ of the quadratic form $H$ depend on the same variables.
For the sake of convenience we rewrite equation (\ref{Heqs0_SvSok}) in the form solved with respect to $U_{xx}$:
\begin{eqnarray} \label{eq1_U2F_SvSok}
U_{xx}=F(U,U_x,u,u_x)
\end{eqnarray}
where $F=a_0U_x+b_0U+h_0R_0$, $R_0=\sqrt{c_0UU_x+r_0U_x^2+s_0U^2}$. The functions $a_0$, $b_0$,\dots,$s_0$ depend on $u,u_x$.

{\bf Remark.} The degenerate case when $F$ is a rational function of $U,U_x$ leads to a contradiction.

Evidently the consistency condition of the equations (\ref{eq1_lin_SvSok}) and (\ref{eq1_U2F_SvSok}) gives rise to the following equation
\begin{eqnarray} \label{eq1_found_F_SvSok}
\left.D_tF-D_xU_t\right|_{(3.1),(3.3),(3.5)}=0.
\end{eqnarray}
By doing all of the differentiations in (\ref{eq1_found_F_SvSok}) and then expressing the appeared variables through the dynamical ones $U,U_x,u,u_x,u_{xx},...$ we obtain an overdetermined equation the searched function $F$ should satisfy to
\begin{eqnarray} \label{eq1_found_F_compl_SvSok}
\fl -3\Bigl(F_{u_xu_x}u_{xx}+u_xU_x+U_xF_{Uu_x}+FF_{U_xu_x}+u_xF_{uu_x}-\sin u\cos u U\Bigr)u_{xxx}-F_{u_xu_xu_x}u^3_{xx}\nonumber\\
\fl -3\Bigl(U_x+U_xF_{Uu_xu_x}+F_{uu_x}+F_{U_xu_x}F_{u_x}+FF_{U_xu_xu_x}+u_xF_{uu_xu_x}\Bigr)u^2_{xx}\nonumber\\
\fl -3\Bigl(FF_{U_xu}+FF_{Uu_x}+u^2_xF_{uuu_x}+F^2F_{U_xU_xu_x}-3\sin u\cos u U_x+\sin u\cos u UF_{U_x}\Bigr.\nonumber\\
\fl \Bigl.+2U_xFF_{U_xUu_x}+U_xF_{U_xu_x}F_U+U_xF_{uU}+U_xF_{U_xU}F_{u_x}+2u_xU_xF_{Uuu_x}+2u_xF+u_xF_{uu}\Bigr.\nonumber\\
\fl \Bigl.+U^2_xF_{UUu_x}+u_xF_{U_xu_x}F_{u}+FF_{U_xU_x}F_{u_x}+2u_xFF_{U_xuu_x}+FF_{U_x}F_{U_xu_x}+3u_xU\sin^2u\Bigr.\nonumber\\
\fl \Bigl.+u_xF_{u_x}F_{U_xu}-u_xU_xF_{U_x}-3u_xU\cos^2u\Bigr)u_{xx}-3u_xU\sin u\cos uF_U-F^3F_{U_xU_xU_x}\nonumber\\
\fl -u^3_xF_{uuu}-3F^2F_{U_xU}-U^3_xF_{UUU}-3u_xFF_uF_{U_xU_x}-3u_xFF_{U_x}F_{U_xu}-u^3_xF_u\\
\fl -3u^2_x\sin u\cos uF_{u_x}+9u_x\sin u\cos uF-12u^3_xU\sin u\cos u-3u_xF^2F_{U_xU_xu}\nonumber\\
\fl -3u^2_xU\cos^2uF_{U_x}+3u^2_xU\sin^2uF_{U_x}-3U_xF^2F_{U_xU_xU}-3U_xu_xF_uF_{U_xU}-3U^2_xFF_{U_xUU}\nonumber\\
\fl -3u^2_xF_uF_{U_xu}-3u^2_xFF_{U_xuu}-3F^2F_{U_xU_x}F_{U_x}-3u_xU^2_xF_{UUu}-3U^2_xF_{U}F_{U_xU}\nonumber\\
\fl -6u_xU_x\sin u\cos uF_{U_x}-3u_xU_xF_{U_xu}F_{U}-6u_xU_xFF_{U_xUu}-3U_xFF_{UU}-3u_xFF_{Uu}\nonumber\\
\fl -9u^2_xU_x\sin^2u+9u^2_xU_x\cos^2u-3u^2_xU_xF_{Uuu}-3U_xFF_{U_x}F_{U_xU}-3U_xFF_UF_{U_xU_x}. \nonumber
\end{eqnarray}
By comparing the coefficients before the product $u_{xx}u_{xxx}$ in (\ref{eq1_found_F_compl_SvSok}) we observe that $F$ is linear with respect to $u_x$:
\begin{eqnarray} \label{eq1_F_via_F1F2_SvSok}
F(U,U_x,u,u_x)=F_1(U,U_x,u)u_x+F_2(U,U_x,u).
\end{eqnarray}
According to (\ref{eq1_U2F_SvSok}) functions $F_1$ and $F_2$ in (\ref{eq1_F_via_F1F2_SvSok}) are as follows
\begin{eqnarray} \label{eq1_F1_SvSok}
F_1(U,U_x,u)=a_1(u)U_x+b_1(u)U+h_1(u)R_1,
\end{eqnarray}
\begin{eqnarray} \label{eq1_F2_SvSok}
F_2(U,U_x,u)=a_2(u)U_x+b_2(u)U+h_2(u)R_1,
\end{eqnarray}
where
\begin{eqnarray} \label{eq1_R0_SvSok}
R_1=\sqrt{c(u)UU_x+r(u)U_x^2+s(u)U^2}.
\end{eqnarray}
Here the functions $a_j(u)$, $b_j(u)$, $h_j(u)$ for $j=1,2$ and $c(u)$, $r(u)$, $s(u)$ are to be found.
Discuss first the choice of the factors $h_1$, $h_2$. We have three possibilities:
\begin{eqnarray*} \label{eq1_poss_h_SvSok}
1) h_1h_2\neq 0, \quad 2) h_1\equiv 0, \quad 3) h_2\equiv 0.
\end{eqnarray*}
It is easily verified that the first two choices lead to a contradiction. In what follows we concentrate on the third possibility.
Let us substitute (\ref{eq1_F_via_F1F2_SvSok})-(\ref{eq1_R0_SvSok}) into (\ref{eq1_found_F_compl_SvSok}), compare the coefficients before the independent variables $u_x,u_{xx},u_{xxx},U,U_x,R_1$ and obtain the following six equations:
\begin{enumerate}
	\item $c(u)=-2a_2(u)r(u);$
	\item $s(u)=-b_2(u)r(u)-\frac{1}{2}c(u)a_2(u);$
	\item $b_1(u)=-a_1(u)a_2(u);$
	\item $h_1(u)c(u)b_2(u)=0;$
	\item $a_1(u)b_2(u)=\sin u\cos u;$
	\item $b'_2(u)-3\sin u\cos u+b_2(u)a_1(u)=0.$
\end{enumerate}
A simple analysis of the equations (i)-(vi) allows one to conclude due to the inequality $\sin u\cos u\neq 0$ that 
\begin{eqnarray*}
\fl 1) \,c(u)=a_2(u)=b_1(u)=0;\\
\fl 2) \,s(u)=-b_2(u)r(u);\\
\fl 3) \,\mbox{none of the functions} \, r(u), b_2(u), s(u), a_1(u) \, \mbox{vanishes identically}.
\end{eqnarray*}
By subtracting equation (v) from (vi) we obtain a differential equation $b'_2(u)=(\sin^2u)'$ for the unknown $b_2(u)$ which implies
\begin{eqnarray*} \label{eq1_b2_SvSok}
b_2(u)=\sin^2u+k_1, \quad k_1=const.
\end{eqnarray*}
Now equation (v) gives rise to
\begin{eqnarray*} \label{eq1_a1_SvSok}
a_1(u)=\frac{\sin u\cos u}{\sin^2u+k_1}.
\end{eqnarray*}
Let us specify the form of the function $F$ by using the computations above and substitute it again into the equation (\ref{eq1_found_F_compl_SvSok}). After comparing the coefficients before $u_xu_{xxx}U_x^2$ we get a differential equation
\begin{eqnarray*} \label{eq1_cond_h1_SvSok}
\fl h'_1(u)(2r(u)\sin^2u+2r(u)k_1)+h_1(u)(4r(u)\sin u\cos u+r'(u)\sin^2u+r'(u)k_1)=0
\end{eqnarray*}
which is easily solved
\begin{eqnarray*} \label{eq1_h1_SvSok}
h_1(u)=\frac{k_2}{(\sin^2u+k_1)\sqrt{r(u)}}.
\end{eqnarray*}
The rest part of the defining equation (\ref{eq1_found_F_compl_SvSok}) is as follows
\begin{eqnarray*} \label{eq1_k2_SvSok}
k_2^2=-(k_1^2+k_1).
\end{eqnarray*}
Let us denote $\lambda:=k_1$ and write down the invariant manifold desired in a final form 
\begin{eqnarray} \label{eq1_final_invmn_SvSok}
\fl U_{xx}=\frac{\sin u\cos u u_x}{\sin^2u+\lambda}U_x+(\sin^2u+\lambda)U+\frac{\sqrt{\lambda(\lambda+1)}u_x}{\sin^2u+\lambda}\sqrt{(\sin^2u+\lambda)U^2-U_x^2}.
\end{eqnarray}
Now we have to find a change of the variables linearizing equation (\ref{eq1_final_invmn_SvSok}). We express the variables $U$, $U_x$ as some quadratic forms of the new variables $\varphi$, $\psi$ chosen is such a way that the function under the square root is a complete square of a combination of the variables $\varphi$ and $\psi$. Assume that
\begin{equation}\label{q1}
U=\alpha\varphi^2+2\beta\varphi\psi+\gamma\psi^2, \quad U_x=a\varphi^2+2b\varphi\psi+c\psi^2.
\end{equation}
Actually we are introducing the vector valued function $\Phi=(\varphi, \psi)^T$ satisfying a linear equation which is just the first equation of the searched Lax pair
\begin{equation}\label{q2}
\Phi_x=A(u,u_x,\dots,\lambda)\Phi.
\end{equation}
Since the equation (\ref{q2}) is defined up to a linear transformation $\Phi\rightarrow S\Phi$ we can by applying the linear transformation bring one of the quadratic forms (\ref{q1}) (say the first one) to the canonical form $U=\varphi^2+\psi^2$. It is easily checked that the expression $sU^2-U_x^2$ with $s=\lambda+\sin^2u$ is a complete square if and only if the following three conditions hold
$$1)ab=0, \quad 2) bc=0, \quad 3) (s-2b^2-ac)^2=(s-a^2)(s-c^2).$$ 
Here we have two choices a) $b=0$ and b) $a=0$, $c=0$. The case a) is not acceptable since it implies $a=c$ and then the map $(\varphi, \psi)\rightarrow (U,U_x)$ defined by (\ref{q1}) has the degenerate Jacobian. In the case b) the third equation yields $s=b^2$. Thus the appropriate change of the variables is as follows:
\begin{eqnarray} \label{eq1_UU1_fipsi_SvSok}
U=\varphi^2+\psi^2, \quad U_x=2\sqrt{\sin^2u+\lambda}\varphi\psi.
\end{eqnarray}
Then equation (\ref{eq1_final_invmn_SvSok}) is reduced to a system of linear equations of the following form
\begin{equation} 
\label{Lax_x_eq1_SvSok}
\left \{
\begin{array}{l}
\varphi_{x}=\frac{1}{2}\left(\sqrt{\sin^2u+\lambda}-\frac{\sqrt{\lambda(\lambda+1)}u_x}{\sin^2u+\lambda}\right)\psi,\\
\psi_{x}=\frac{1}{2}\left(\sqrt{\sin^2u+\lambda}+\frac{\sqrt{\lambda(\lambda+1)}u_x}{\sin^2u+\lambda}\right)\varphi.
\end{array} 
\right.
\end{equation}
Equation (\ref{eq1_lin_SvSok}) becomes:
\begin{equation} 
\label{Lax_t_eq1_SvSok}
\left \{
\begin{array}{l}
\varphi_{t}=\frac{1}{2}\frac{\sqrt{\lambda(\lambda+1)}u_{xx}}{\sqrt{\sin^2u+\lambda}}\varphi+\frac{1}{4}\left(\frac{\sqrt{\lambda(\lambda+1)}\Bigl(u_x(\sin^2u-2\lambda-u_x^2)-2u_{xxx}\Bigr)}{\sin^2u+\lambda}\right.\\
\qquad \left.-\frac{(\sin^2u+\lambda)(\sin^2u-2\lambda-u_x^2)-2\sin u\cos uu_{xx}}{\sqrt{\sin^2u+\lambda}}\right)\psi,\\
\psi_{t}=-\frac{1}{2}\frac{\sqrt{\lambda(\lambda+1)}u_{xx}}{\sqrt{\sin^2u+\lambda}}\psi-\frac{1}{4}\left(\frac{\sqrt{\lambda(\lambda+1)}\Bigl(u_x(\sin^2u-2\lambda-u_x^2)-2u_{xxx}\Bigr)}{\sin^2u+\lambda}\right.\\
\qquad \left.+\frac{(\sin^2u+\lambda)(\sin^2u-2\lambda-u_x^2)-2\sin u\cos uu_{xx}}{\sqrt{\sin^2u+\lambda}}\right)\varphi.
\end{array} 
\right.
\end{equation}
It can be checked that the system of equations (\ref{Lax_x_eq1_SvSok}), (\ref{Lax_t_eq1_SvSok}) define the Lax pair to the equation (\ref{eq1_SvSok}).

\subsection{Equation (\ref{eq2_SvSok})}

In a similar way we can construct the Lax pair for the equation (\ref{eq2_SvSok}). First we look for the nonlinear invariant manifold to the linearized equation:
\begin{eqnarray}\label{eq2_lin}
U_t=U_{xxx}-\frac{3u_xu_{xx}}{1+u_x^2}U_{xx}+\frac{3}{2}\left(u_x^2+\frac{(u_x^2-1)u_{xx}^2}{(u_x^2+1)^2}\right)U_x.
\end{eqnarray}
We find that equation (\ref{eq2_lin}) is consistent with the surface (evidently it is  defined by a quadratic form):
\begin{eqnarray}\label{eq2_quadr_form}
U_{xx}=\frac{u_xu_{xx}}{1+u_x^2}U_x+\frac{1}{\lambda}(1+u_x^2)U+\frac{\sqrt{1+\lambda}u_x}{\lambda}\sqrt{(1+u_x^2)U^2-\lambda U_x^2}.
\end{eqnarray}
The change of the variables:
\begin{eqnarray}\label{eq2_fipsi}
U=\varphi^2+\psi^2, \quad U_x=\frac{2\sqrt{1+u_x^2}}{\sqrt{\lambda}}\varphi\psi
\end{eqnarray}
brings the system (\ref{eq2_lin}), (\ref{eq2_quadr_form}) to a  Lax pair for (\ref{eq2_SvSok}):
\begin{equation} 
\label{Lax_x_eq2_SvSok}
\left \{
\begin{array}{l}
\varphi_{x}=\frac{1}{2\sqrt{\lambda}}\left(\sqrt{1+u^2_x}-u_x\sqrt{1+\lambda}\right)\psi,\\
\psi_{x}=\frac{1}{2\sqrt{\lambda}}\left(\sqrt{1+u^2_x}+u_x\sqrt{1+\lambda}\right)\varphi.
\end{array} 
\right.
\end{equation}

\begin{equation} 
\label{Lax_t_eq2_SvSok}
\left \{
\begin{array}{l}
\varphi_{t}=\frac{1}{4\lambda^\frac{3}{2}}\left(\frac{(1+u_x^2)^2(\lambda u_x^2+2)+2u_xu_{xxx}\lambda(1+u_x^2)-u_{xx}^2\lambda(3u_x^2+1)}{(1+u_x^2)^{\frac{3}{2}}}\right.\\
\quad \left.-\frac{((1+u_x^2)(\lambda u_x^3+2u_x+2u_{xxx}\lambda)-3u_{xx}^2u_x\lambda)\sqrt{1+\lambda}}{(1+u_x^2)}\right)\psi+\frac{1}{2}\frac{\sqrt{1+\lambda}u_{xx}}{\sqrt{1+u_x^2}\lambda}\varphi\\
\psi_{t}=\frac{1}{4\lambda^\frac{3}{2}}\left(\frac{(1+u_x^2)^2(\lambda u_x^2+2)+2u_xu_{xxx}\lambda(1+u_x^2)-u_{xx}^2\lambda(3u_x^2+1)}{(1+u_x^2)^{\frac{3}{2}}}\right.\\
\quad \left.+\frac{((1+u_x^2)(\lambda u_x^3+2u_x+2u_{xxx}\lambda)-3u_{xx}^2u_x\lambda)\sqrt{1+\lambda}}{(1+u_x^2)}\right)\varphi-\frac{1}{2}\frac{\sqrt{1+\lambda}u_{xx}}{\sqrt{1+u_x^2}\lambda}\psi\\
\end{array} 
\right.
\end{equation}

\section{Formal diagonalization of the found Lax pairs and the local conservation laws}
In this section we find formal asymptotic expansions for the eigenfunctions of the Lax pairs (\ref{Lax_x_eq1_SvSok}), (\ref{Lax_t_eq1_SvSok}) and (\ref{Lax_x_eq2_SvSok}), (\ref{Lax_t_eq2_SvSok}) around the singular points of the spectral parameter $\lambda$. From these asymptotic expansions we deduce the conservation laws for the equations (\ref{eq1_SvSok}) and (\ref{eq2_SvSok}). It is generally accepted that the Lax pair which  produces the infinite series of the conserved densities is certainly the true Lax pair. 

Let us begin with the system (\ref{Lax_x_eq1_SvSok}), (\ref{Lax_t_eq1_SvSok}). By the linear transformation $\Phi=\tilde{T}Y$ where $\Phi=(\varphi,\psi)^T$ and $\tilde{T}=\left( \begin{array} {cc}  1& -1\\  1& 1 \end{array}\right)$  we reduce (\ref{Lax_x_eq1_SvSok}), (\ref{Lax_t_eq1_SvSok}) to the following form 
\begin{equation} \label{eq1_diag_AG}
Y_x=AY, \quad Y_t=GY.
\end{equation}
Here the matrices $A, G$ are given by
\begin{equation*}
A=\left( 
\begin{array} {cc}  \frac{1}{2}\sqrt{\sin^2u +\lambda}& -\frac{1}{2}\frac{u_x \sqrt{\lambda(\lambda+1)}}{\sin^2u +\lambda}\\
  \frac{1}{2}\frac{u_x \sqrt{\lambda(\lambda+1)}}{\sin^2u +\lambda}&  -\frac{1}{2}\sqrt{\sin^2u +\lambda}
\end{array}\right),
\end{equation*}
\begin{eqnarray*}
\fl G=\left( 
\begin{array} {cc} \frac{1}{4}\frac{2u_{xx}\sin u \cos u+(\sin^2u +\lambda)(u_x^2-\sin^2u +2\lambda)}{\sqrt{\sin^2u +\lambda}} & 
-\frac{\sqrt{\lambda(\lambda+1)}}{4\sqrt{\sin^2u +\lambda}}\left(2u_{xx}+\frac{2u_{xxx}+u_x(u_x^2-\sin^2u +2\lambda)}{\sqrt{\sin^2u +\lambda}}\right)
\\ \frac{\sqrt{\lambda(\lambda+1)}}{4\sqrt{\sin^2u +\lambda}}\left(\frac{2u_{xxx}+u_x(u_x^2-\sin^2u +2\lambda)}{\sqrt{\sin^2u +\lambda}}-2u_{xx}\right) & -\frac{1}{4}\frac{2u_{xx}\sin u \cos u+(\sin^2u +\lambda)(u_x^2-\sin^2u +2\lambda)}{\sqrt{\sin^2u +\lambda}}
\end{array}\right).
\end{eqnarray*}
The matrix $A$ has singularities at the points $\lambda=0$, $\lambda=-1$ and $\lambda=\infty.$ Let us expand $A$ around $\lambda=\infty:$ 
\begin{equation}\label{eq1_A_series}
A=\sum^{\infty}_{j=-1}A_j\lambda^{-j/2}, \qquad A_{-1}=\frac{1}{2}\left( \begin{array} {cc}  1& 0\\ 0 & -1 \end{array}\right).
\end{equation}
According to the general theory (see \cite{Wasow}, \cite{Wils}, \cite{DrS}) we look for a formal change of the variables $\Psi=TY$ transforming (\ref{eq1_diag_AG}) to the form 
\begin{equation} \label{eq1_diag_hS}
\Psi_x=h\Psi, \quad \Psi_t=S\Psi, 
\end{equation}
where $T$, $h$ and $S$ are formal power series:
\begin{equation}\label{eq1_hT_series}
T=\sum^{\infty}_{j=0}T_j\lambda^{-j/2} , \qquad h=\sum^{\infty}_{j=-1}h_j\lambda^{-j/2}, \qquad S=\sum^{\infty}_{j=-3}S_j\lambda^{-j/2}.
\end{equation}
The matrices $h_j$ are assumed to be diagonal. Setting 
$T_0=1$ and assuming that for $\forall \, i\geq 1$ all diagonal entries of $T_i$ vanish we find the coefficients of the series $T$ and $h$ from the equation $T_x=AT-Th$ . By comparing the coefficients in 
\begin{eqnarray}\label{eq1_series_found_Th}
\fl \qquad \sum^{\infty}_{j=1}D_x(T_j)\lambda^{-j/2}=\sum^{\infty}_{j=-1}A_j\lambda^{-j/2}\sum^{\infty}_{j=0}T_j\lambda^{-j/2}-\sum^{\infty}_{j=0}T_j\lambda^{-j/2}\sum^{\infty}_{j=-1}h_j\lambda^{-j/2}
\end{eqnarray}
we obtain a sequences of the equations for defining $T_j,h_j:$
\begin{equation} 
\label{eq1_found_Th}
\begin{array}{l}
h_{-1}=A_{-1},\\
\left[A_{-1},T_1\right]-h_0=-A_0,\\
\left[A_{-1},T_2\right]-h_1=D_x(T_1)-A_0T_1-A_1+T_1h_0,\\
\cdots
\end{array} 
\end{equation}
The system of the equations (\ref{eq1_found_Th}) is consecutively solved. Omitting the computations we give only the answers
\begin{eqnarray*}
\fl h=\left( \begin{array} {cc}  \frac{1}{2}& 0\\ 0 & -\frac{1}{2} \end{array}\right)\lambda^{\frac{1}{2}}+\left( \begin{array} {cc}  -\frac{1}{4}(u_x^2-\sin^2u)& 0\\ 0 & \frac{1}{4}(u_x^2-\sin^2u) \end{array}\right)\lambda^{-\frac{1}{2}}+ \\
\qquad \left( \begin{array} {cc}  \frac{1}{4}u_xu_{xx}& 0\\ 0 & \frac{1}{4}u_xu_{xx} \end{array}\right)\lambda^{-1} +\left( \begin{array} {cc}  h_{3,11}& 0\\ 0 & -h_{3,11} \end{array}\right)\lambda^{-\frac{3}{2}}+\dots,
\end{eqnarray*}

where $h_{3,11}=\frac{5}{8}u_x^2\sin^2u-\frac{1}{4}u_x^2-\frac{1}{16}u_x^4-\frac{1}{4}u_xu_{xxx}-\frac{1}{16}\sin^4u$

\begin{eqnarray*}
\fl T=\left(\begin{array} {cc}  1& 0\\ 0 & 1 \end{array}\right)+\left( \begin{array} {cc}  0& \frac{1}{2}u_x\\ \frac{1}{2}u_x & 0 \end{array}\right)\lambda^{-\frac{1}{2}}+\left( \begin{array} {cc}  0& \frac{1}{2}u_{xx}\\ -\frac{1}{2}u_{xx} & 0 \end{array}\right)\lambda^{-1}\\
\fl +\left( \begin{array} {cc}  0& \frac{1}{2}u_{xxx}+\frac{1}{4}u_{x}+\frac{1}{8}u_x^3-\frac{3}{4}u_x\sin^2u\\ \frac{1}{2}u_{xxx}+\frac{1}{4}u_{x}+\frac{1}{8}u_x^3-\frac{3}{4}u_x\sin^2u & 0 \end{array}\right)\lambda^{-\frac{3}{2}}+\dots
\end{eqnarray*}

For known $T$ and $h$ the series $S$ is defined as follows
\begin{equation}\label{eq1_S_series}
S=T^{-1}GT-T^{-1}T_t=\sum^{\infty}_{j=-3}S_j\lambda^{-j/2} 
\end{equation}
due to the expansion of $G$ around $\lambda=\infty$: $G=\sum^{\infty}_{j=-3}G_j\lambda^{-j/2}.$ It can be verified that the coefficients of the series $S$ are diagonal:

\begin{eqnarray*}
\fl S=\left(\begin{array} {cc}  \frac{1}{2}& 0\\ 0 & -\frac{1}{2} \end{array}\right)\lambda^{\frac{3}{2}}+\left(\begin{array} {cc}  S_{1,11} & 0\\ 0 & -S_{1,11} \end{array}\right)\lambda^{-\frac{1}{2}}+\\
\qquad \left(\begin{array} {cc}  S_{2,11} & 0\\ 0 & S_{2,11} \end{array}\right)\lambda^{-1}+\left(\begin{array} {cc}  S_{3,11} & 0\\ 0 & -S_{3,11} \end{array}\right)\lambda^{-\frac{3}{2}}+\dots, 
\end{eqnarray*}
where 
\begin{eqnarray*}
 \fl \qquad S_{1,11}=\frac{1}{16}u_x^2(14\sin^2u-3u_x^2-4)-\frac{1}{2}u_xu_{xxx}+\frac{1}{4}u_{xx}(u_{xx}+\sin 2u)-\frac{3}{16}\sin^4u,\\
 \fl \qquad S_{2,11}=\frac{1}{4}u_xu_{xxxx}-\frac{3}{8}u_xu_{xx}\sin^2u+\frac{3}{8}u_x^3u_{xx}-\frac{3}{8}u_x^3\sin 2u,\\
 \fl \qquad S_{3,11}=-\frac{1}{4}u_xu_{xxxxx}+\frac{1}{4}u_{xx}u_{xxxx}-\frac{1}{4}u_{xxx}^2+\frac{1}{8}(13\sin^2u-4-5u_x^2)u_xu_{xxx}+ \\
 \qquad \frac{1}{8}(2-3u_x^2-5\sin^2u)u_{xx}^2+\frac{1}{16}\sin^6u+\frac{1}{8}(7u_x^2-\sin^2u)u_{xx}\sin 2u-\\ 
 \qquad \frac{1}{16}u_x^6+\frac{1}{4}(3\cos^2u-\frac{1}{4}\sin^2u-1)u_x^4-\frac{1}{16}(23\sin^2u-12)u_x^2\sin^2u.
\end{eqnarray*}
The consistency condition of the system (\ref{eq1_diag_hS}) 
\begin{equation} \label{eq1_consv_laws_formula}
D_th=D_xS
\end{equation}
shows that $h$ and $S$ are generating functions for the local conservation laws. 
Equation (\ref{eq1_consv_laws_formula}) generates infinite series of the conservation laws for the equation (\ref{eq1_SvSok}). We give two of them in an explicit form
\begin{eqnarray*} 
\label{eq1_cons_laws}
\begin{array}{l}
 \fl D_t\left(u_x^2-\sin^2u\right)=\\
D_x\left(2u_xu_{xxx}-u_{xx}^2-u_{xx}\sin 2u+\frac{3}{4}u_x^4+u_x^2-\frac{7}{2}u_x^2\sin^2u+\frac{3}{4}\sin^4u\right),\\
 \fl D_t\left(4u_{xx}^2+10u_x^2\sin^2u-4u_x^2-u_x^4-\sin^4u\right)=\\
D_x\left(8u_{xx}u_{xxxx}-4u_{xxx}^2-4u_x(u_x^2-5\sin^2u+2)u_{xxx}+\sin^6u+\right.\\
\quad \left.4(3u_x^2-4\sin^2u+1)u_{xx}^2-2\sin 2u(\sin^2u+5u_x^2)u_{xx}+\right.\\
\quad \left.(11\sin^2u-4-u_x^2)u_x^4-(23\sin^2u-12)u_x^2\sin^2u\right).\\
\end{array} 
\end{eqnarray*}
In a similar way we can investigate the Lax pair (\ref{Lax_x_eq2_SvSok}), (\ref{Lax_t_eq2_SvSok}). Here we give only two local conservation laws to the equation (\ref{eq2_SvSok}), evaluated by the same method of formal series
\begin{eqnarray*} 
\label{eq2_cons_laws}
\begin{array}{l}
 \fl D_t\Bigl(\frac{u_{xx}}{u_x\sqrt{u_x^2+1}}\Bigr)=D_x\Bigl(\frac{u_{xxxx}}{u_x\sqrt{u_x^2+1}}-\frac{3u_{xx}u_{xxx}}{(u_x^2+1)^{3/2}}+\frac{3(u_x^2-1)u_{xx}^3}{2u_x(u_x^2+1)^{5/2}}+\frac{3u_xu_{xx}}{2\sqrt{u_x^2+1}}\Bigr),\\
 
 \fl D_t\left(\frac{u_{xx}^2}{u_x^2+1}-u_x^2\right)=\\
\fl \qquad D_x\left(\frac{2u_{xx}u_{xxxx}}{1+u_x^2}-\frac{u_{xxx}^2}{1+u_x^2}-\left(\frac{4u_xu_{xx}^2}{(1+u_x^2)^2}+2u_x\right)u_{xxx}+\frac{(9u_x^2-11)u_{xx}^4}{4(1+u_x^2)^3}+\frac{(11u_x^2+2)u_{xx}^2}{2(1+u_x^2)}-\frac{3}{4}u_x^4\right).\\
\end{array} 
\end{eqnarray*}

\section*{Conclusions}

In the article we discussed a direct algorithm for constructing the Lax pairs to given integrable equations. The essence of the algorithm is in computing the quadratic form which defines an invariant manifold for the linearized equation. Let us briefly comment  the main steps of the method:
\begin{itemize}
\item linearize the given equation and simplfy if possible the linearized equation by a properly chosen linear substitution;
\item find a quadratic form, consistent with the linearized equation. As a result one finds a nonlinear Lax pair for the given equation;
\item look for a transformation reducing the obtained nonlinear Lax pair to a linear one. Usually this transformation is defined by another quadratic form.
\end{itemize}

We illustrated the efficiency of the algorithm by applying it to the well studied models (\ref{kdv_eq}) and (\ref{modVch}). By using the algorithm we also obtained the Lax pairs for the equations (\ref{eq1_SvSok}), (\ref{eq2_SvSok}). In the last section 4 we demonstrated through constructing formal eigenfunctions and   the conservation laws that the found Lax pairs  are not fake.

\section*{Acknowledgments}

The authors gratefully acknowledge financial support from a Russian Science Foundation grant (project 15-11-20007).

\end{document}